\documentclass[12pt]{article}

\pdfoutput=1
\usepackage{graphicx}
\usepackage{amssymb}
\usepackage{amsmath}
\usepackage{color}
\usepackage{cite}
\usepackage{subfigure}
\usepackage{bm}
\usepackage[colorlinks=true,linkcolor=blue,citecolor=blue]{hyperref}
\usepackage{soul}

\begin{document}

\newcommand{\beq}{\begin{equation}}
\newcommand{\eeq}{\end{equation}}


\begin{titlepage}

\begin{flushright}
\end{flushright}

\vskip 2cm

\begin{center}

{\Large \bf
Stochastic gravitational wave background \vspace{2mm} \\ from early dark energy
}

\vspace{1cm}

Naoya Kitajima$^{\,a,b}$
and
Tomo Takahashi$^{\,c}$ \\

\vskip 1.0cm

{\it
$^a$Frontier Research Institute for Interdisciplinary Sciences, Tohoku University, Sendai 980-8578, Japan\\[2mm]
$^b$Department of Physics, Tohoku University, Sendai 980-8578, Japan\\
$^c$Department of Physics, Saga University, Saga 840-8502, Japan
}

\vskip 1.0cm

\begin{abstract}
We study the production of stochastic gravitational wave background from early dark energy (EDE) model.
It is caused by resonant amplification of scalar field fluctuations, which easily takes place for typical EDE potential based on the string axion or $\alpha$-attractor model.
The resultant spectrum of gravitational wave background is computed by performing 3D lattice simulations. We show that, specifically in some class of generalized $\alpha$-attractor EDE model, a significant amount of gravitational waves can be produced via tachyonic instability with a peak around femto-Hz frequency range. Models predicting such gravitational waves can be constrained by the cosmic microwave background observations.
\end{abstract}

\end{center}

\end{titlepage}

\newpage

\vspace{1cm}

\section{Introduction} \label{sec:intro}

Recent observational discrepancy of the Hubble constant $H_0$, known as the Hubble tension,  where the values of $H_0$ derived from direct and indirect measurements are inconsistent at almost 5$\sigma$ level (see e.g., \cite{DiValentino:2021izs,Perivolaropoulos:2021jda} for a review of the current status of the tension), motivates us to consider the new physics beyond the $\Lambda$CDM model.
One of such models suggested to resolve the Hubble tension is an early dark energy (EDE) scenario \cite{Poulin:2018cxd} in which a scalar field fractionally contributes to the total energy density at around recombination epoch. As a potential for the early dark energy, the axion type one is assumed in many works. Indeed axion is one of the representative example of the new physics which is originally motivated by the particle physics \cite{Peccei:1977hh,Peccei:1977ur,Weinberg:1977ma,Wilczek:1977pj} and is also predicted by string theory \cite{Svrcek:2006yi,Arvanitaki:2009fg,Cicoli:2012sz}.
Therefore axion can be a plausible candidate for the EDE because it has generally a large initial misalignment amplitude while the mass is extremely small \cite{Kamionkowski:2014zda,Karwal:2016vyq}.
An extremely small mass is required because the axion should be frozen until the recombination epoch to be regarded as EDE
and a large initial amplitude is also needed because the axion should give a sizable contribution to the total energy density in order to affect the expansion of the Universe, which can reduce the sound horizon to increase the value of $H_0$ when fitted to
cosmic microwave background (CMB) data such as Planck \cite{Planck:2018nkj}.

Another motivation for the axion EDE is the cosmic birefringence, recently reported in \cite{Minami:2020odp} using the Planck data. In general, the background axion EDE field can induce the rotation of the linear polarization angle of the CMB photon and hence it can explain the reported isotropic cosmic birefringence \cite{Fujita:2020ecn,Choi:2021aze,Nakatsuka:2022epj,Murai:2022zur,Eskilt:2023nxm}.

In this paper, we discuss the consequences of the axion EDE other than the Hubble tension and the cosmic birefringence. In particular, we show that the stochastic gravitational wave (GW) background is predicted from the axion EDE scenario through the 2nd order scalar field fluctuations.
Actually the scalar field fluctuations can be resonantly amplified through the self-interaction or interaction with other field as in the case of the preheating after inflation  \cite{Kofman:1994rk,Kofman:1997yn}. For the case of the preheating, a significant amount of GW can be produced with the spectrum peaked typically at GHz range \cite{Khlebnikov:1997di,Easther:2006gt,Easther:2006vd,Garcia-Bellido:2007nns,Garcia-Bellido:2007fiu,Figueroa:2017vfa,Adshead:2018doq}.
By the same mechanism, string axion can also induce stochastic GWs in lower frequency bands depending on the axion mass \cite{Soda:2017dsu,Kitajima:2018zco,Machado:2018nqk,Machado:2019xuc} and axion dark matter scenario typically predicts GWs with nHz frequency range \cite{Ratzinger:2020koh,Namba:2020kij,Kitajima:2020rpm,Ratzinger:2020oct,Co:2021rhi,Madge:2021abk}.
Here in this paper, we argue that the GWs are inevitably generated in the axion EDE scenario for some particular parameter choice \cite{Johnson:2008se}, which can be regarded as a unique cosmological signature of the axion EDE\footnote{
Actually GW production from the axion EDE model is studied in \cite{Weiner:2020sxn} where the explosive gauge-field production is assumed, which is a different scenario from the one we consider in this paper.
}.
To this end, we assume two types of potential for the EDE, which has been motivated by string theories and show that the GW spectrum is peaked around the frequency of $f\sim 10^{-15}~{\rm Hz}$ which 
is a new source of GWs at the frequency region.

The structure of this paper is as follows. In the next section, we describe the axionic early dark energy model we consider in this paper and discuss the evolution of the scalar field and the spectrum of density fluctuations obtained from lattice simulation. Then in Section~\ref{sec:gw}, we discuss the production of GWs in the scenario. The GW spectrum generated in our model is presented. The final section is devoted to conclusions and discussions.

\section{Early dark energy and parametric resonance} \label{sec:res}

\begin{figure}[tp]
\centering
\includegraphics [width = 9cm, clip]{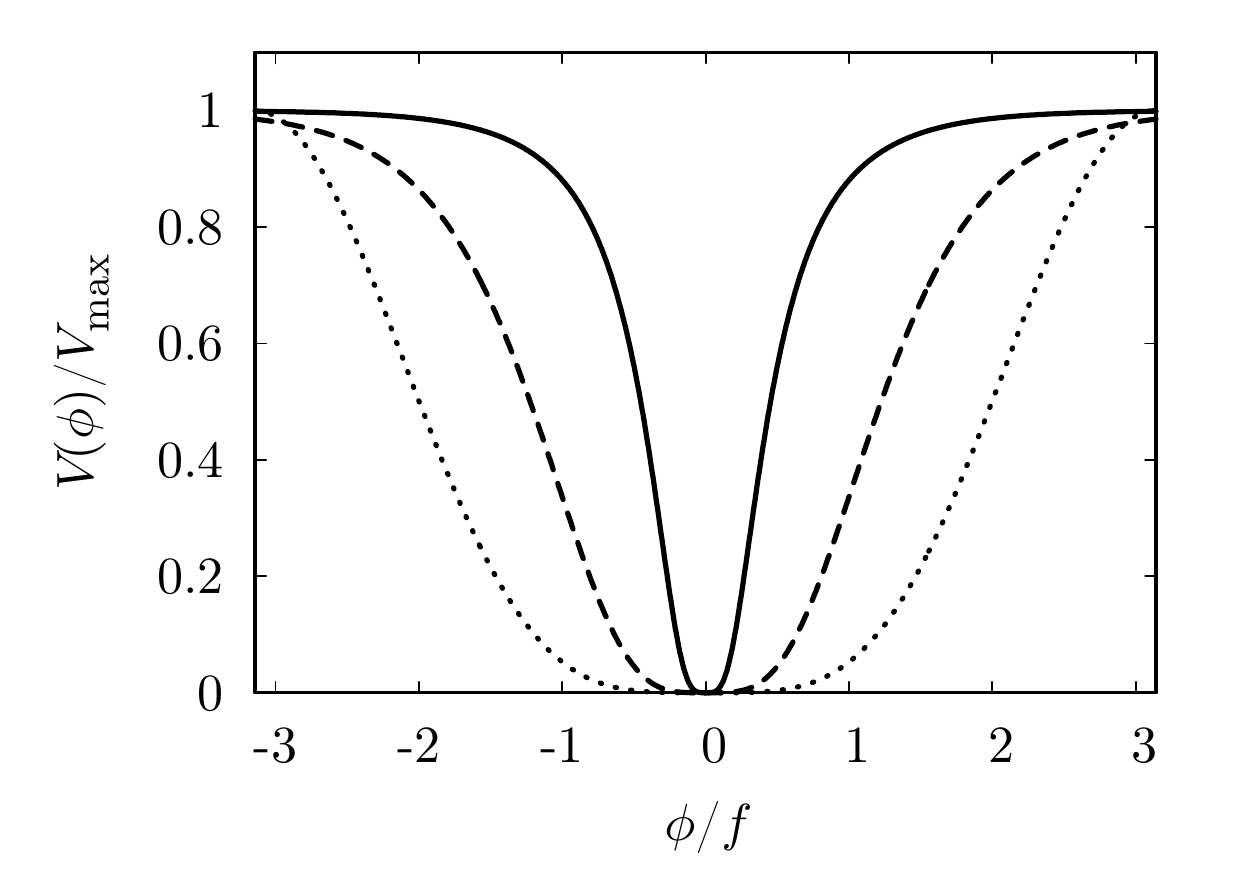}
\caption{
The shape of the potential for the axion EDE type with $n=2$ (dotted), the $\alpha$-attractor-type with $(p,q,r,\beta)=(2,0,0,0)$ (dashed) and the generalized $\alpha$-attractor-type with $(p,q,r,\beta)=(2,2,1,10)$ (solid). For the $\alpha$-attractor cases, $f$ corresponds to $f=\sqrt{6 \alpha} M_P$. 
}
\label{fig:potential}
\end{figure}

Let us consider the model with the following axion-like EDE potential,
\begin{align}
V(\phi) = m^2 f^2 \left[1-\cos\bigg(\frac{\phi}{f}\bigg)\right]^n,
\end{align}
where $n$ is an integer, $f$ is the decay constant and $m$ is the energy scale which determines the onset of the field oscillation.
We also consider a generalized $\alpha$-attractor type potential \cite{LinaresCedeno:2019bgo,Braglia:2020bym} as follows:
\begin{align}
V(\phi) = 
\frac{V_0 (1+\beta)^{2r} \tanh^{2p}(\phi/\sqrt{6\alpha}M_P)}{[1+\beta \tanh^q (\phi/\sqrt{6\alpha}M_P)]^{2r}} \,,
\end{align}
where $p$, $q$ and $r$ are (positive) integers and $\beta$ is a numerical constant. $V_0$ corresponds to the potential value at $\phi \to \infty$. $\alpha$ is a parameter which controls the energy scale of the scalar field.
Both of the above potentials can be approximated by $m^2f^2(\phi/f)^{2n}/2^n$ (axionlike), and $V_0(1+\beta)^{2r}(\phi/\sqrt{6\alpha}M_P)^{2p}$ ($\alpha$-attractor) for small $\phi$ limit.
The shape of the potential is illustrated in Figure \ref{fig:potential}. Assuming the flat Friedmann‐Lema\^{i}tre‐Robertson-Walker (FLRW) background metric, the field equation is given by
\begin{align} \label{eq:scalarEoM}
\ddot{\phi}+3H\dot{\phi}-\frac{\nabla^2 \phi}{a^2}+\frac{\partial V}{\partial\phi}=0,
\end{align}
where the overdot represents the derivative with respect to the cosmic time, $a$ is the scale factor and $H=\dot{a}/a$ is the Hubble parameter.
The Hubble parameter is determined by the background radiation/matter mixed fluid.
The scalar field is initially placed far from the minimum (i.e. misaligned) and almost frozen due to the Hubble friction term. During this period, the energy density of the scalar field is kept constant and thus it behaves like a dark energy with equation of state $w \simeq -1$. The scalar field starts to roll down the potential when the Hubble friction becomes inefficient against the potential gradient and then it oscillates around the minimum of the potential. In this phase, it is no longer regarded as dark energy, and its energy density dilutes faster than that of matter component 
in order to be consistent with the observed $\Lambda$CDM cosmology
when the EDE field starts to oscillate at early times.
Therefore, the $n=1$ case is in general not compatible with the early dark energy model especially in the context of the Hubble tension\footnote{
Considering the interaction between the axion and dark photon, the axion abundance can be significantly suppressed due to the tachyonic production of dark photons. The suppression factor can be typically as small as $10^{-2}$ -- $10^{-3}$ \cite{Kitajima:2017peg,Agrawal:2018vin,Kitajima:2020rpm}. It may open up a possibility of the axion EDE with $n=1$.
}.

The oscillation of the scalar field is highly anharmonic in the case with $n\geq 2$. In other words the self-interaction always exists. In this case, the nonzero (inhomogeneous) mode can be excited as a consequence of tachyonic instability or parametric resonance instability. It shows exponential amplifications of the field fluctuation and the inhomogeneous mode easily dominates over the homogeneous coherent oscillation mode.
In particular, the case with $n=2$ is a special one because it approaches conformally invariant model ($V = (\lambda/4) \phi^4$) in the small field limit. In this case, the resonant amplification persistently occurs until the backreaction turns on \cite{Greene:1997fu}. 
In the subsequent sections, we show the dynamics of this amplification process by numerical lattice simulations.

In what follows, we focus on the following specific cases:
\begin{itemize}
    \item (a) Axion EDE-type potential with $n=2$,
    \item (b) Simple $\alpha$-attractor-type potential with $(p,q,r,\beta)=(2,0,0,0)$,
    \item (c) Generalized $\alpha$-attractor-type potential with $(p,q,r,\beta) = (2,2,1,10)$, 
    \item (d) Generalized $\alpha$-attractor-type potential with $(p,q,r,\beta) = (3,2,1.5,10)$.
\end{itemize}

\subsection{Lattice simulation} \label{subsec:lattice}

\begin{figure}[tp]
\centering
\subfigure[Axion EDE]{
\includegraphics [width = 7.5cm, clip]{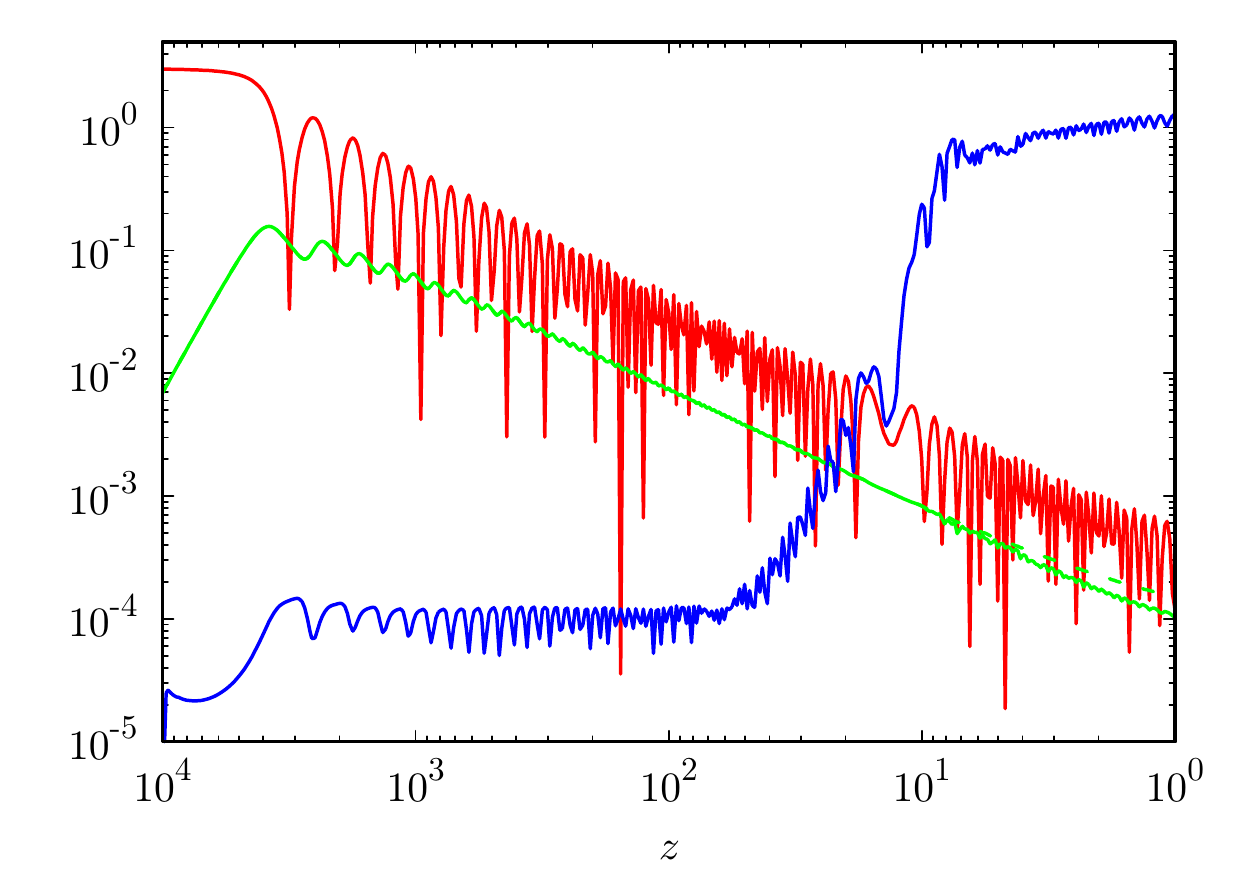}
\label{subfig:evovle_a}
}
\subfigure[$\alpha$-attractor EDE]{
\includegraphics [width = 7.5cm, clip]{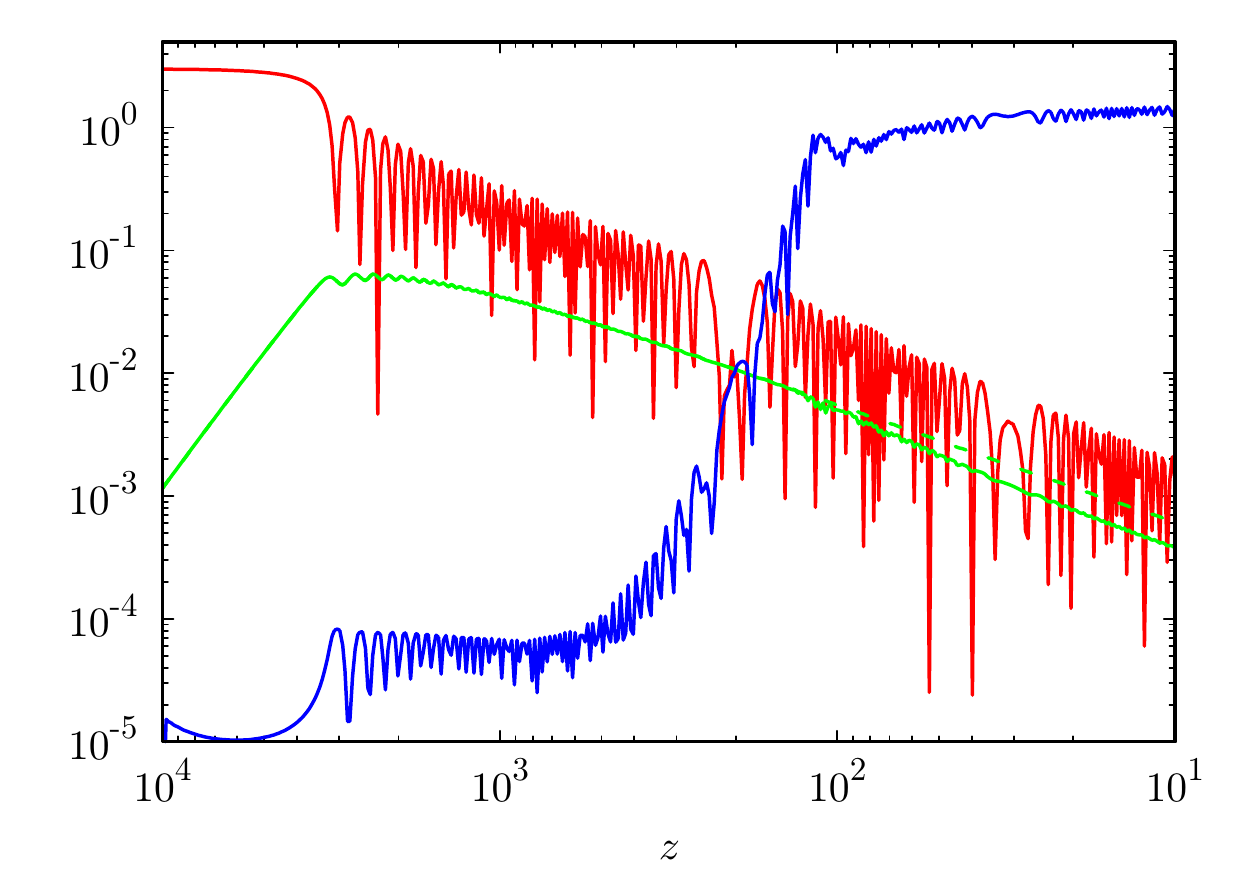}
\label{subfig:evolve_b}
}
\subfigure[Generalized $\alpha$-attractor ($p=2$)]{
\includegraphics [width = 7.5cm, clip]{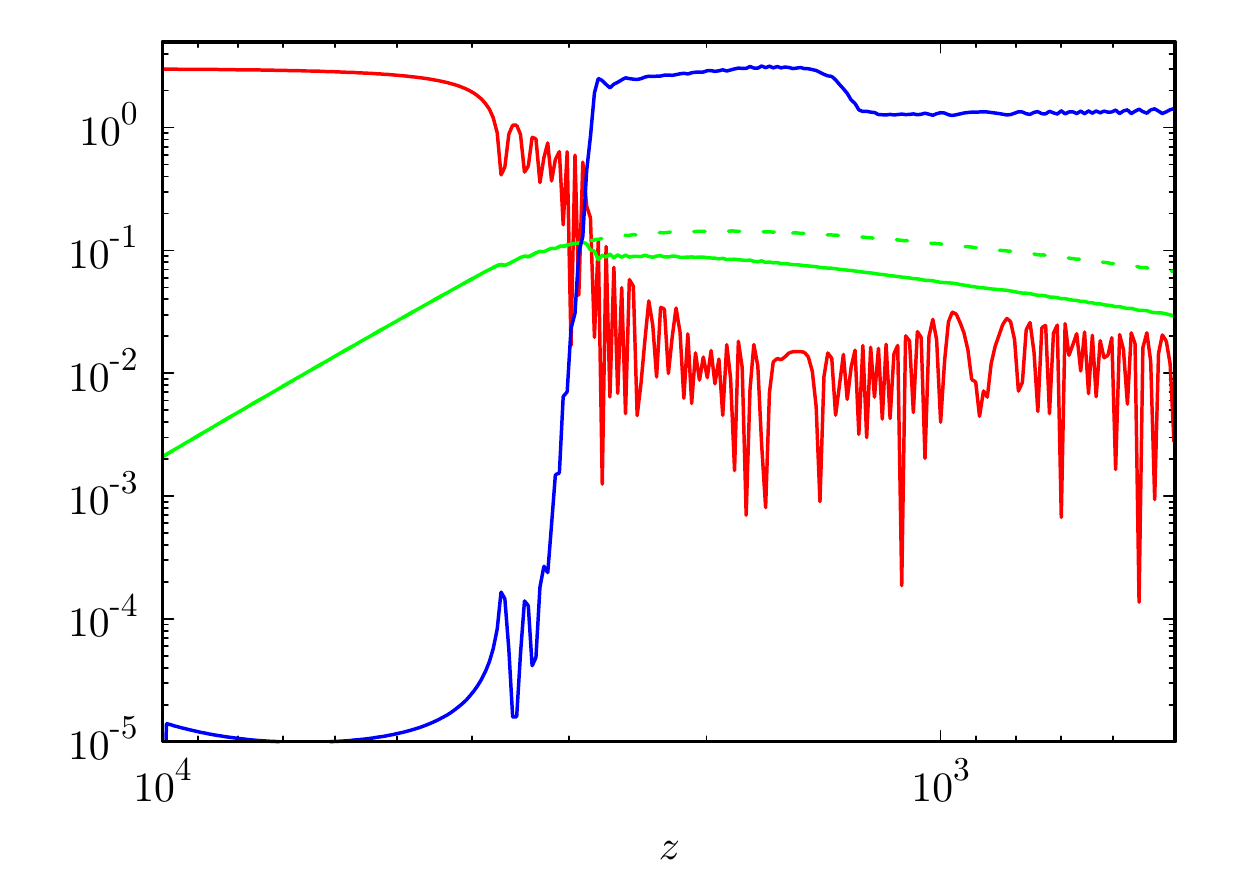}
\label{subfig:evolve_c}
}
\subfigure[Generalized $\alpha$-attractor ($p=3$)]{
\includegraphics [width = 7.5cm, clip]{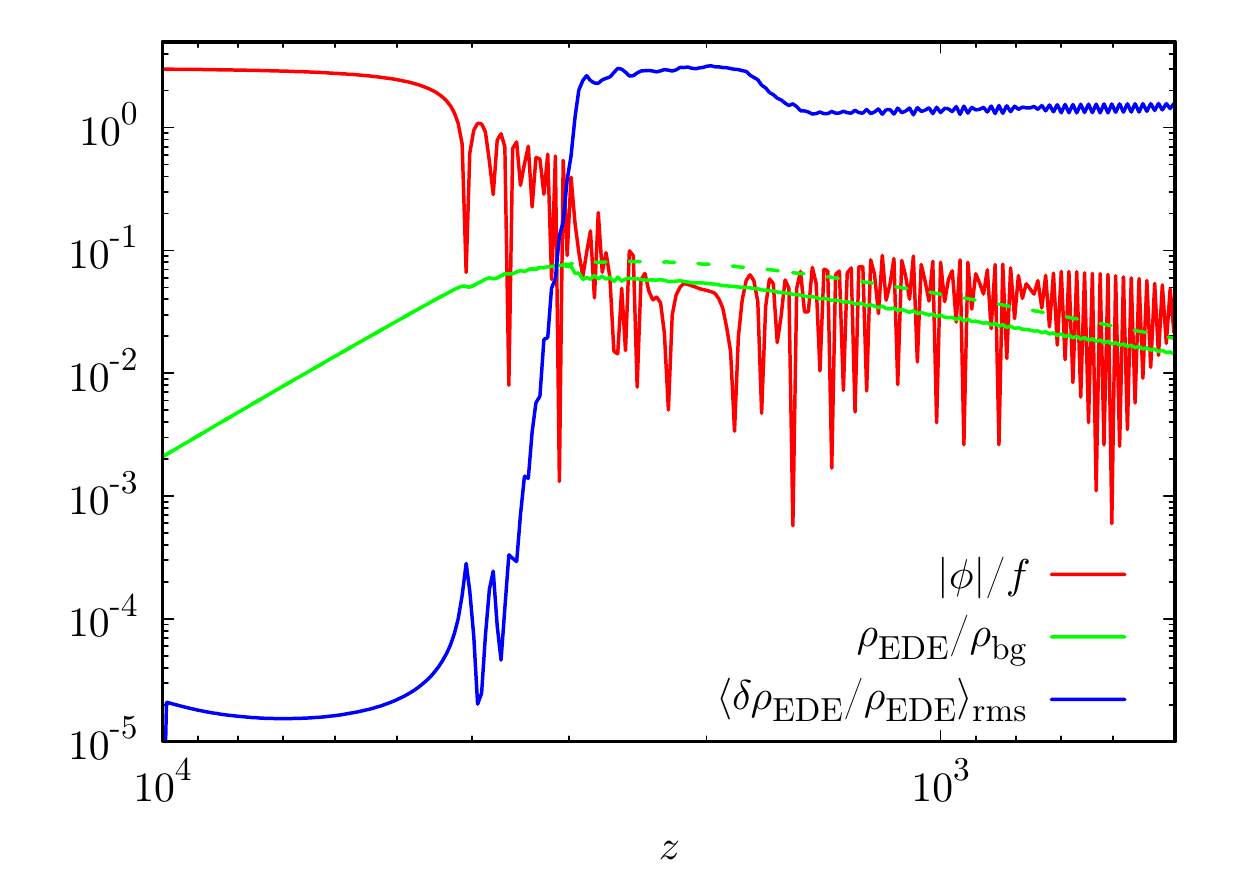}
\label{subfig:evovle_d}
}
\caption{
Time evolution of the zero mode oscillation (red), energy fraction of EDE (green) and root-mean-square of density fluctuation (blue) of EDE. The horizontal axis represents the redshift.
The dashed green line corresponds to the energy fraction of EDE including homogeneous component (without backreaction).
}
\label{fig:evolve}
\end{figure}

Because the scalar field fluctuation (i.e. the nonzero mode) grows exponentially due to the tachyonic/parametric resonance instability, it soon becomes comparable to the homogeneous mode and the growth is saturated at that time. Then, the produced nonzero mode start to backreact to the homogeneous mode and the dynamics enters the nonlinear regime. Thus, in order to correctly follow the
full dynamics of the system, we need to directly solve the evolution equation (\ref{eq:scalarEoM}) with discretized lattice space where the spatial derivative is replaced with the finite difference. The setup of our simulation for each case is summarized in Table~\ref{tab:setup}.
The initial field value for the homogeneous mode is $\phi_i/f = 3$ in all cases ($f=\sqrt{6\alpha}M_P$ for the $\alpha$-attractor cases) and we put random Gaussian field on each grid point with the standard deviation $\delta\phi = 10^{-5}$ as the initial field fluctuations.
The choices of other parameters in the potential have also been listed in Table~\ref{tab:setup}. With the parameter values assumed here, the EDE can give a sizable energy density fraction at around the matter-radiation equality. Indeed a lot of analysis using observational data such as CMB have been performed to find the favored values for the EDE parameters (see, e.g., \cite{Poulin:2023lkg} and the references therein). We may take the best-fit values from those analysis to choose the values of the EDE parameters, however, the preferred ones can depend on the data set adopted and the setup in the analysis. We should also mention that the nonlinear effect discussed below could affect the evolution of EDE to change the fit to the data.  
In the light of these considerations, we just assume the values listed in Table~\ref{tab:setup} as examples. We show the time evolution of the EDE field from our simulations in Fig.~\ref{fig:evolve}.
The red line shows the time evolution of the scalar field, which exhibits the coherent oscillation and the blue line represents the root-mean-square of the density fluctuation of EDE, $\delta\rho_{\rm EDE}/\rho_{\rm EDE}$. 
The green line corresponds to the energy fraction of EDE to the background energy density, $\rho_{\rm EDE}/(\rho_r+\rho_m)$ with $\rho_r$ and $\rho_m$ being that of radiation and matter,  respectively. The dashed green line corresponds to the energy fraction of EDE for the homogeneously oscillating case without the nonlinear effect (backreaction).
Interestingly, by comparing the energy density with and without the nonlinear effect, one can clearly see that the backreaction modifies the evolution of the EDE energy density, which may affect the constraint on EDE from cosmological data such as CMB. Although this would be an important issue, we here focus on the production of GWs through the 2nd order scalar field fluctuations.

\begin{table}
\begin{center}
\begin{tabular}{c c c c c c c c}
\hline
~  & grid points & initial box size & $z$-range & $f$ & $m$ & $\alpha$ & $V_0$  \\
\hline
(a) & $512^3$ & $0.8\pi H_{\rm eq}^{-1}$ & $10^4$\,--\,$1$ & $0.1 M_P$ & $5H_{\rm eq}$ & - & - \\
(b) & $512^3$ & $0.4\pi H_{\rm eq}^{-1}$ & $10^4$\,--\,$10$ & - & - & $10^{-4}$ & $0.17 H_{\rm eq}^2 M_P^2$ \\
(c) & $512^3$ & $0.04\pi H_{\rm eq}^{-1}$ & $10^4$\,--\,$500$ & - & - & $10^{-5}$ & $0.3 H_{\rm eq}^2 M_P^2$ \\
(d) & $512^3$ & $0.04\pi H_{\rm eq}^{-1}$ &$10^4$\,--\,$500$ & - & - & $10^{-5}$ & $0.3 H_{\rm eq}^2 M_P^2$ \\
\hline
\end{tabular}
\end{center}
\caption{Setup of our lattice simulations. Here $H_{\rm eq}$ is the Hubble parameter at the time of matter-radiation equality.}
\label{tab:setup}
\end{table}

For the axion EDE case (case (a)), the axion starts to oscillate at the redshift $z_{\rm osc} \sim 4000$ (which corresponds to $z_c$, the redshift at which the density fraction of EDE is maximized),  but the growth of scalar field fluctuations starts well after the onset of the oscillation ($z \sim 60$) as also shown by the linear analysis in \cite{Smith:2019ihp}. The growth stops when the density contrast becomes ${\cal O}(1)$ at $z \sim 10$.
For a simple $\alpha$-attractor model (case (b)), the growth starts earlier (at $z \sim 600$) and saturates at $z \sim 100$. Note that in the above two cases (a) and (b), the resonant amplification starts after the last scattering surface of CMB ($z \sim 1000$) and hence it has no chance to affect the CMB observations such as the B-mode signal due to the gravitational waves. 

On the other hand, one can see that efficient amplification happens in the cases (c) and (d). It is due to the tachyonic instability soon after the commencement of the oscillation\footnote{
The linear analysis of the field fluctuation from the tachyonic instability with plateau type potential (like a generalized $\alpha$-attractor model in our case) has been studied in \cite{Soda:2017dsu,Kitajima:2018zco,Fukunaga:2019unq}. 
}.  In this case, the growth rate is large (compared with the above two cases) and the fluctuation becomes ${\cal O}(1)$ after several oscillations, well before the last scattering surface. It should be noted that the efficient amplification occurs even for $p=3$ case (case (d)).

\begin{figure}[tp]
\centering
\subfigure[Axion EDE]{
\includegraphics [width = 7.5cm, clip]{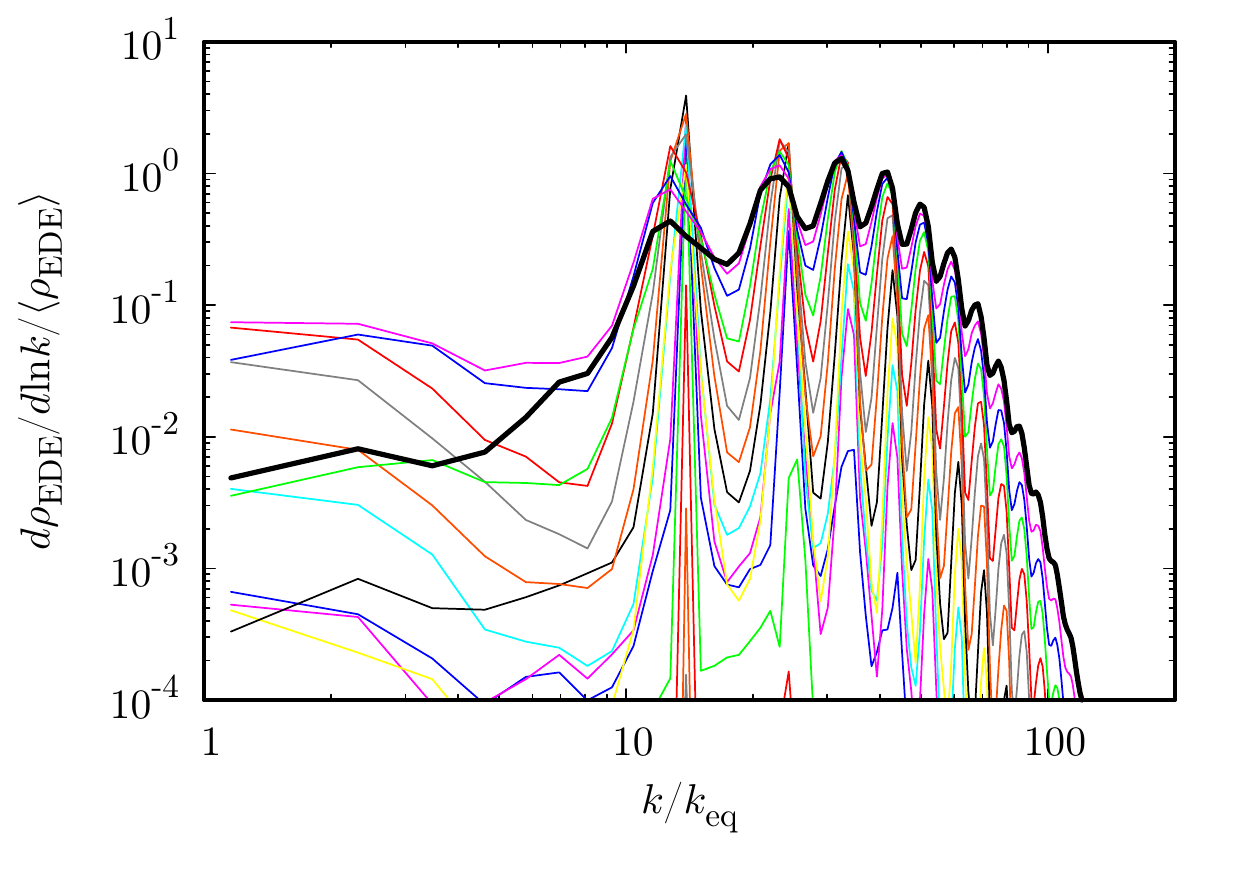}
\label{subfig:spectrum_a}
}
\subfigure[$\alpha$-attractor EDE]{
\includegraphics [width = 7.5cm, clip]{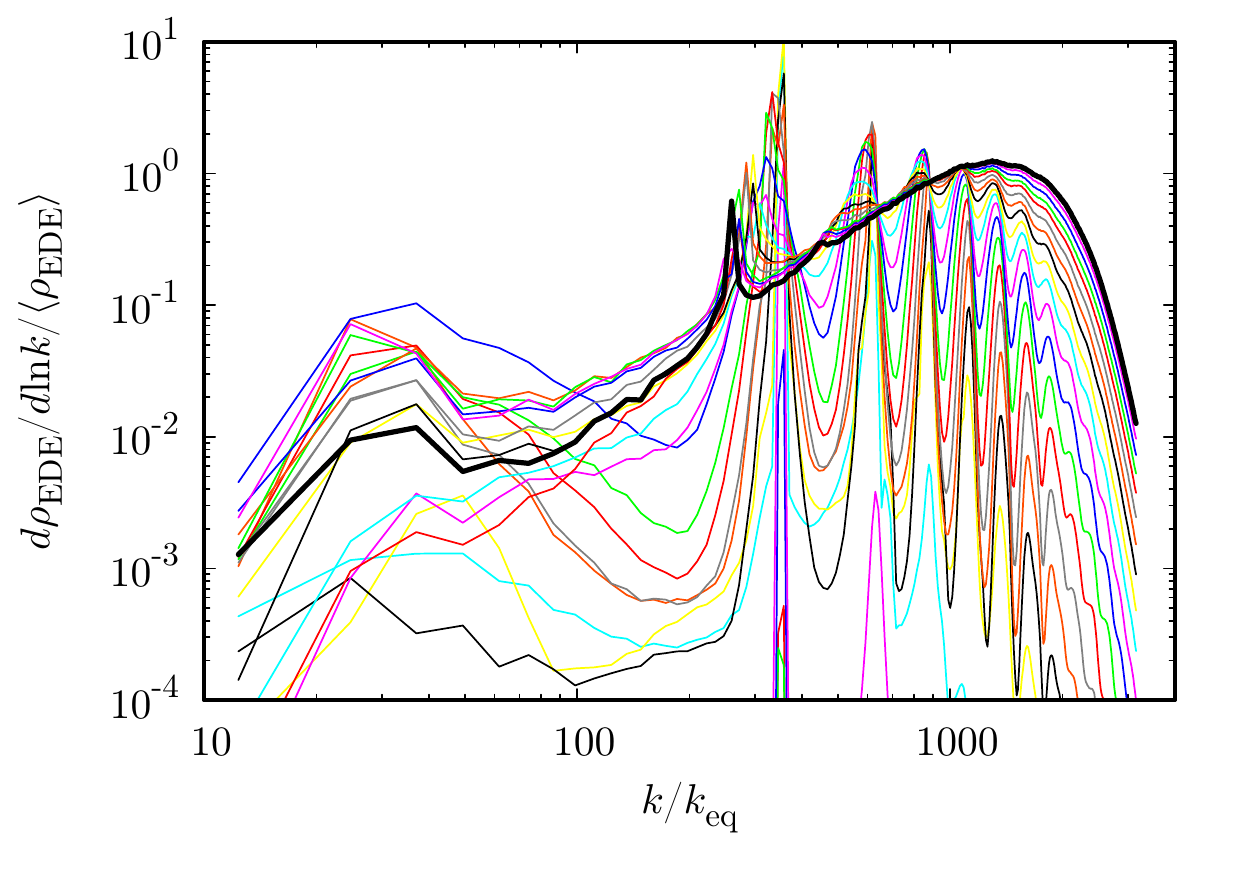}
\label{subfig:spectrum_b}
}
\subfigure[Generalized $\alpha$-attractor ($p=2$)]{
\includegraphics [width = 7.5cm, clip]{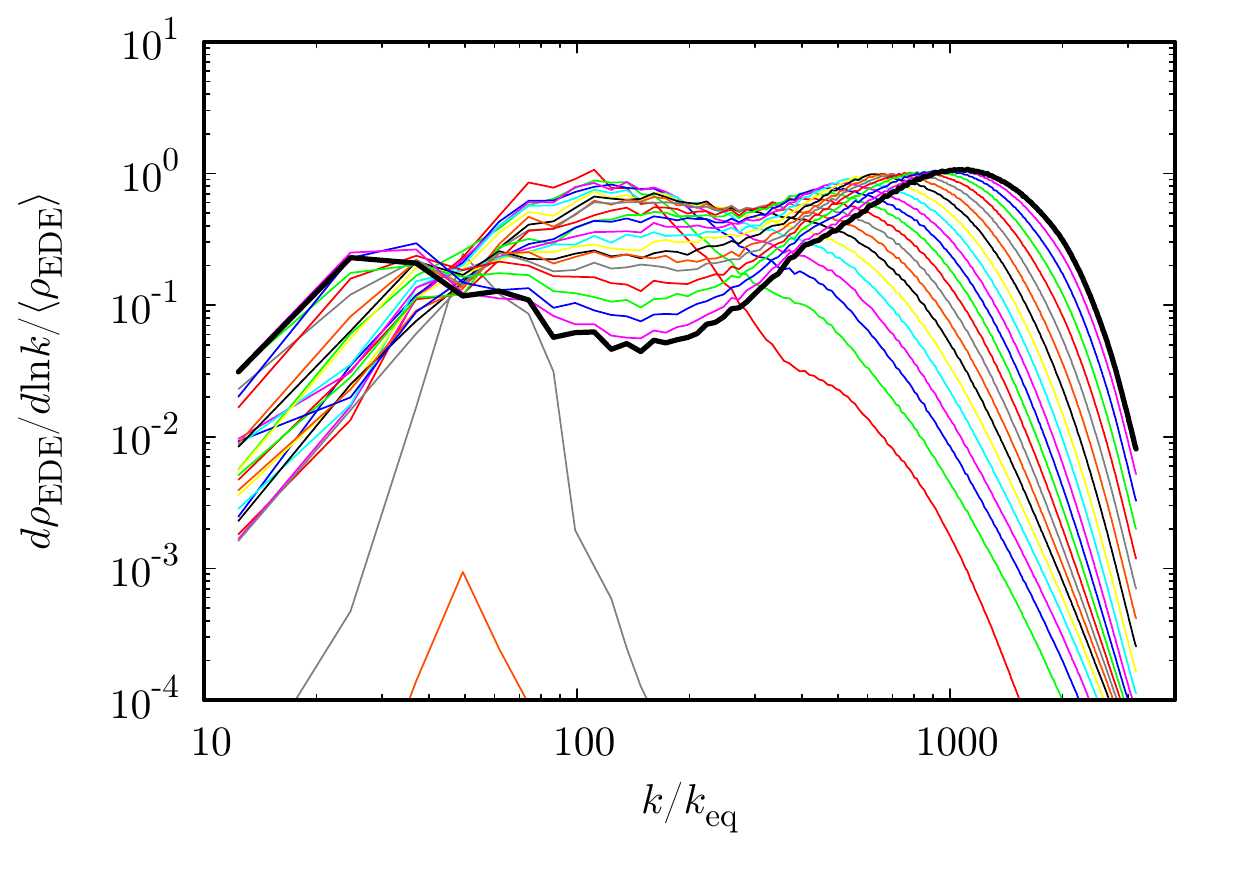}
\label{subfig:spectrum_c}
}
\subfigure[Generalized $\alpha$-attractor ($p=3$)]{
\includegraphics [width = 7.5cm, clip]{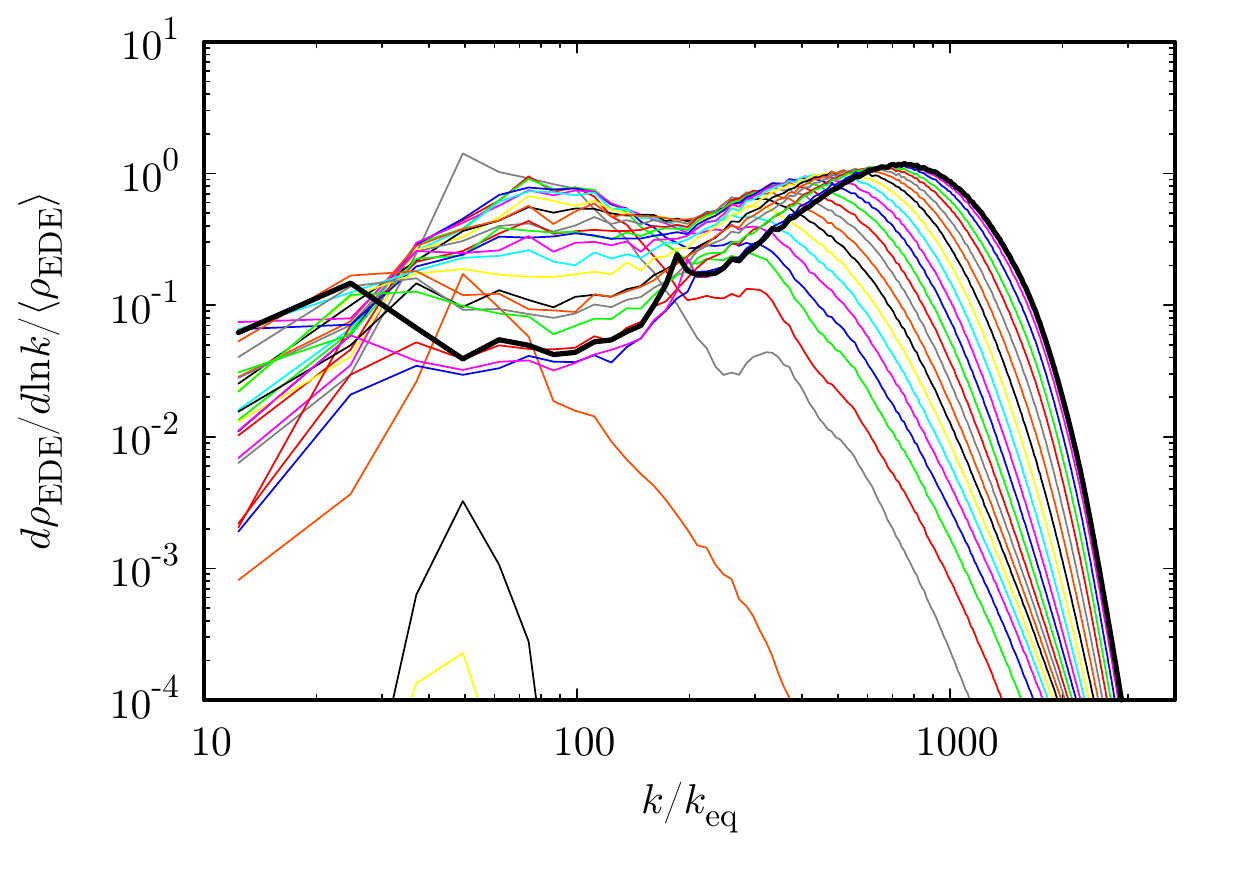}
\label{subfig:spectrum_d}
}
\caption{
Time evolution of the spectrum of the density fluctuation. The time evolves from bottom to top and the thick black line corresponds to the spectrum at the redshift $z=1$ (top-left), $10$ (top-right), and $100$ (bottom panels). The horizontal axis is the wavenumber normalized by the horizon scale at the matter-radiation equality, $k_{\rm eq} = a_{\rm eq}H_{\rm eq}$. 
}
\label{fig:spectrum}
\end{figure}

Fig.~\ref{fig:spectrum} shows the evolution of the spectrum of energy density of the EDE as a function of wavenumber, which is normalized by $k_{\rm eq} = a_{\rm eq} H_{\rm eq}$, corresponding to the horizon scale at the matter-radiation equality. The time evolves from bottom to top and the thick black line corresponds to the spectrum at redshift $z=10$ (top-left), $z=100$ (top-right) and $z=1000$ (bottom panels). 
In the top two figures \ref{subfig:spectrum_a} and \ref{subfig:spectrum_b} (axion EDE and simple $\alpha$-attractor EDE models), one can see that the growth occurs in multiple narrow bands and the spectrum has sharp peaks in the early stages.
It is a typical behaviour of the narrow resonance (see \cite{Kitajima:2018zco} for some detailed discussion in a similar setup). After the growth is saturated, the spectrum is smoothed through the momentum transfer due to the rescattering processes. 
In the bottom two panels \ref{subfig:spectrum_c} and \ref{subfig:spectrum_d} (generalized $\alpha$-attractor models), the growth is due to the tachyonic instability and thus the low-$k$ modes are broadly enhanced in a short period of time.  See \cite{Kitajima:2018zco} for some detailed discussion regarding the growth due to the narrow resonance and the tachyonic instability in a similar setup. Then, the nonlinear processes smooth out the spectrum.
Note that the spectrum has jagged shape in the top left figures (case (a)). It is because the momentum transfer is not completed within the simulation time.  This is in contrast to other cases where the spectral shape is smoothed due to the completion of the momentum transfer. Thus, the nonlinear process can be imprinted in the temporal shape of the spectrum.

\section{Gravitational wave production \label{sec:gw}}

\begin{figure}[tp]
\centering
\subfigure[Axion EDE]{
\includegraphics [width = 7.5cm, clip]{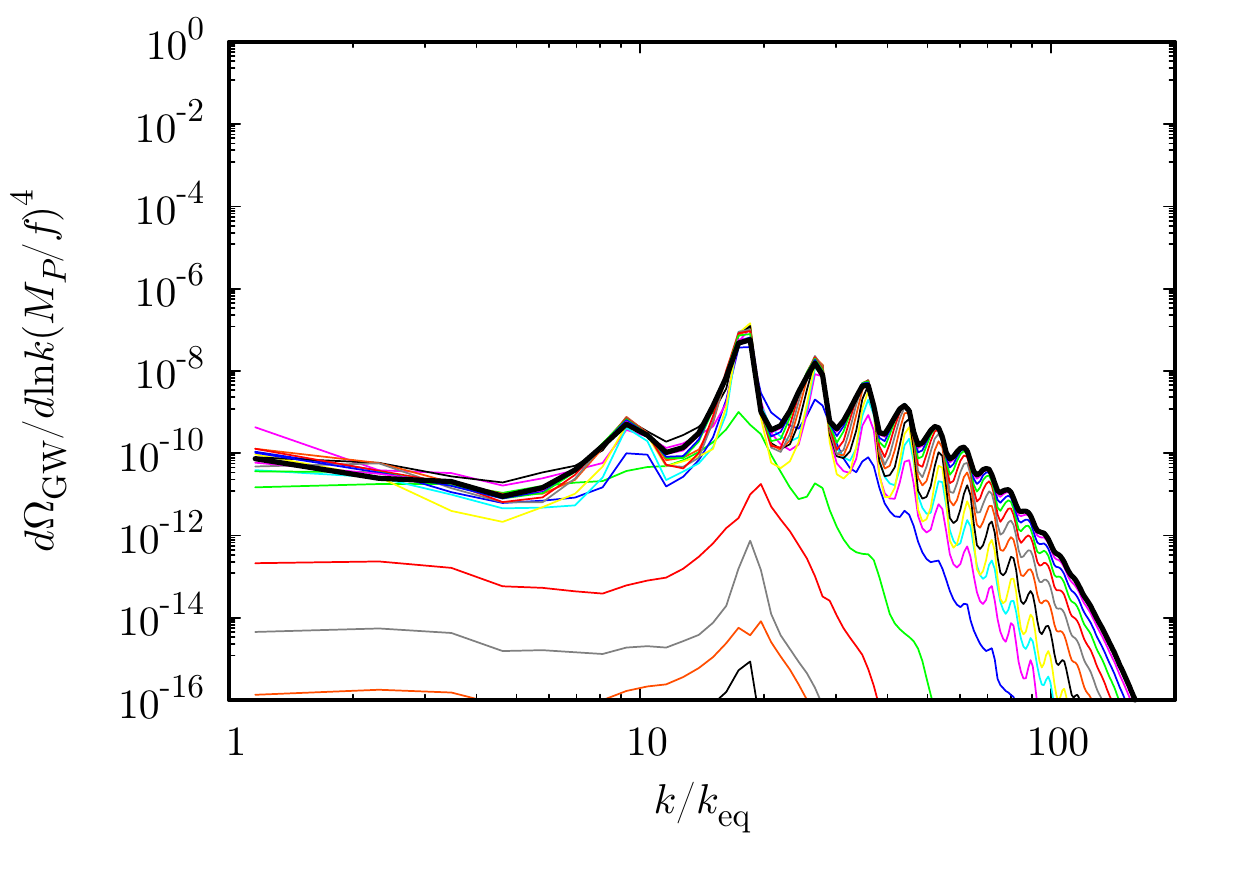}
\label{subfig:OmegaGW_a}
}
\subfigure[$\alpha$-attractor EDE]{
\includegraphics [width = 7.5cm, clip]{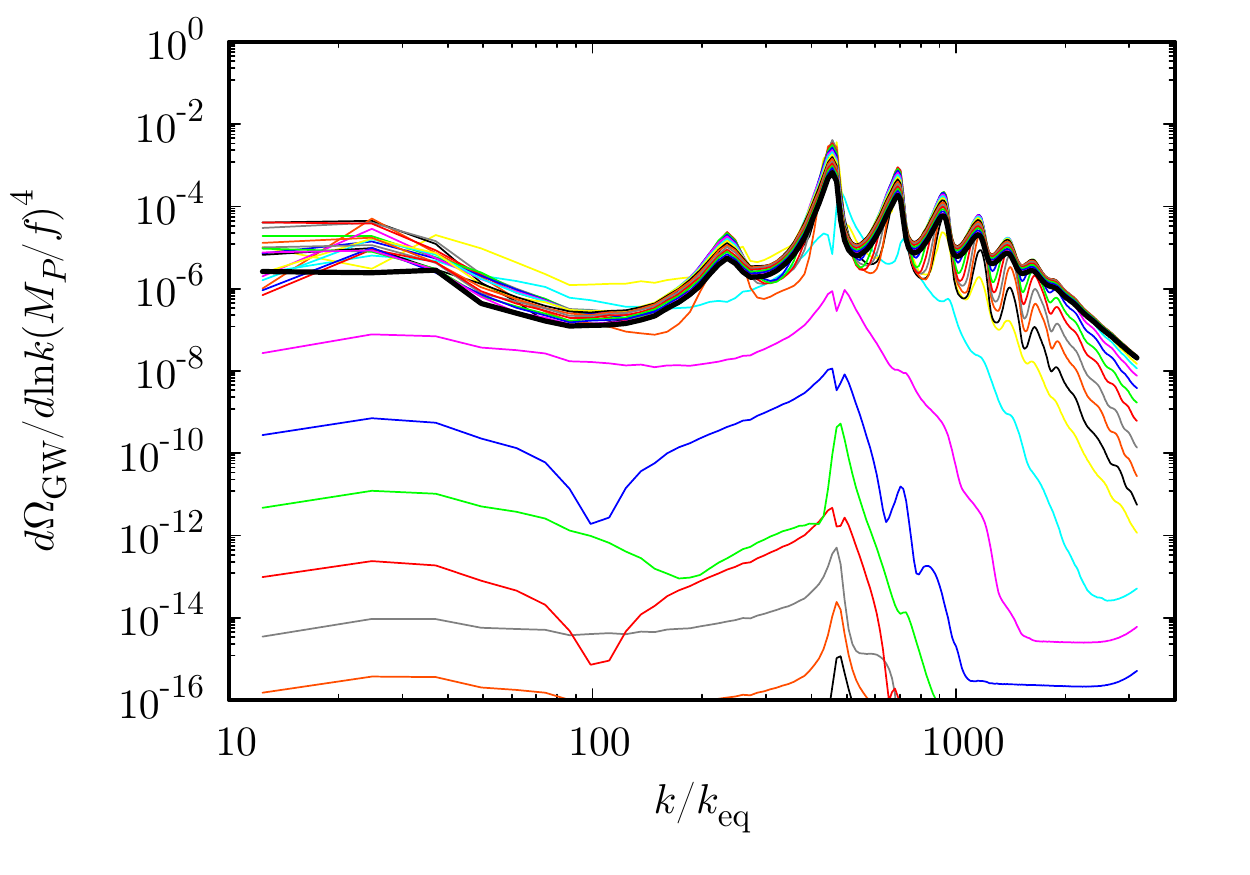}
\label{subfig:OmegaGW_b}
}
\subfigure[Generalized $\alpha$-attractor ($p=2$)]{
\includegraphics [width = 7.5cm, clip]{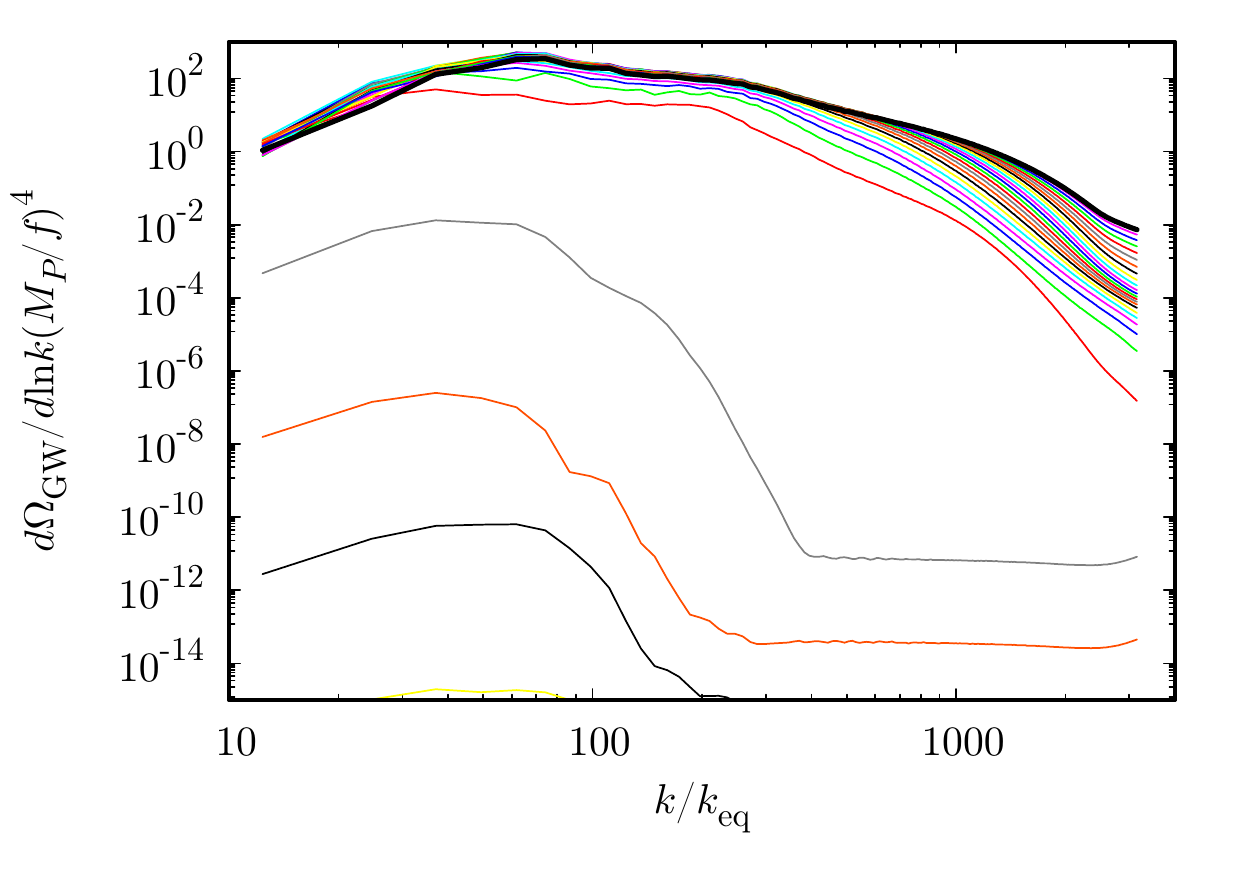}
\label{subfig:OmegaGW_c}
}
\subfigure[Generalized $\alpha$-attractor ($p=3$)]{
\includegraphics [width = 7.5cm, clip]{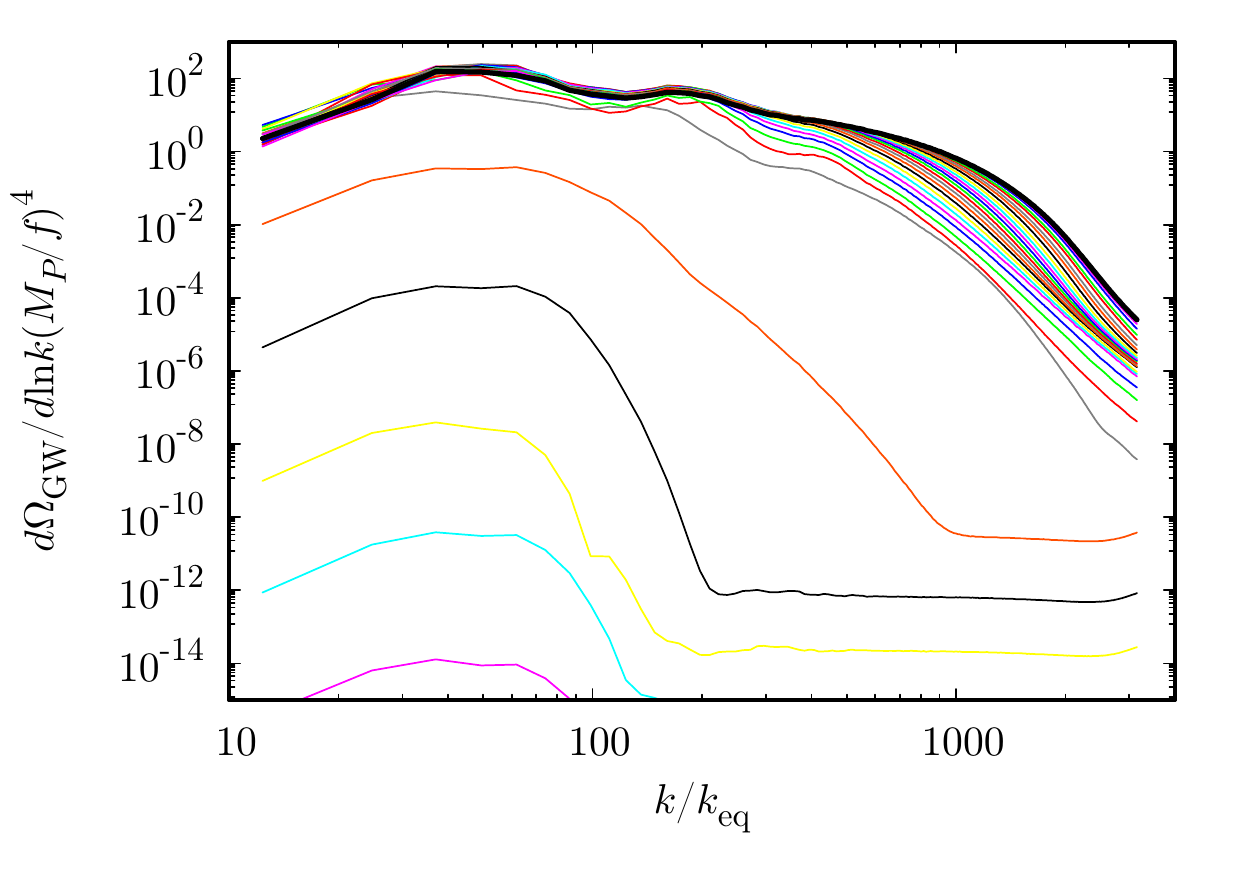}
\label{subfig:OmegaGW_d}
}
\caption{
Time evolution of the spectrum of gravitational waves. The time evolves from bottom to top and the thick black line corresponds to the final time, $z=1$ (top-left), $10$ (top-right), $100$ (bottom panels). The horizontal axis is the wave number normalized by the horizon scale at the matter-radiation equality. We have taken the initial value $\phi_i/f=3$. 
}
\label{fig:OmegaGW}
\end{figure}

\begin{figure}[tp]
\centering
\includegraphics [width = 9cm, clip]{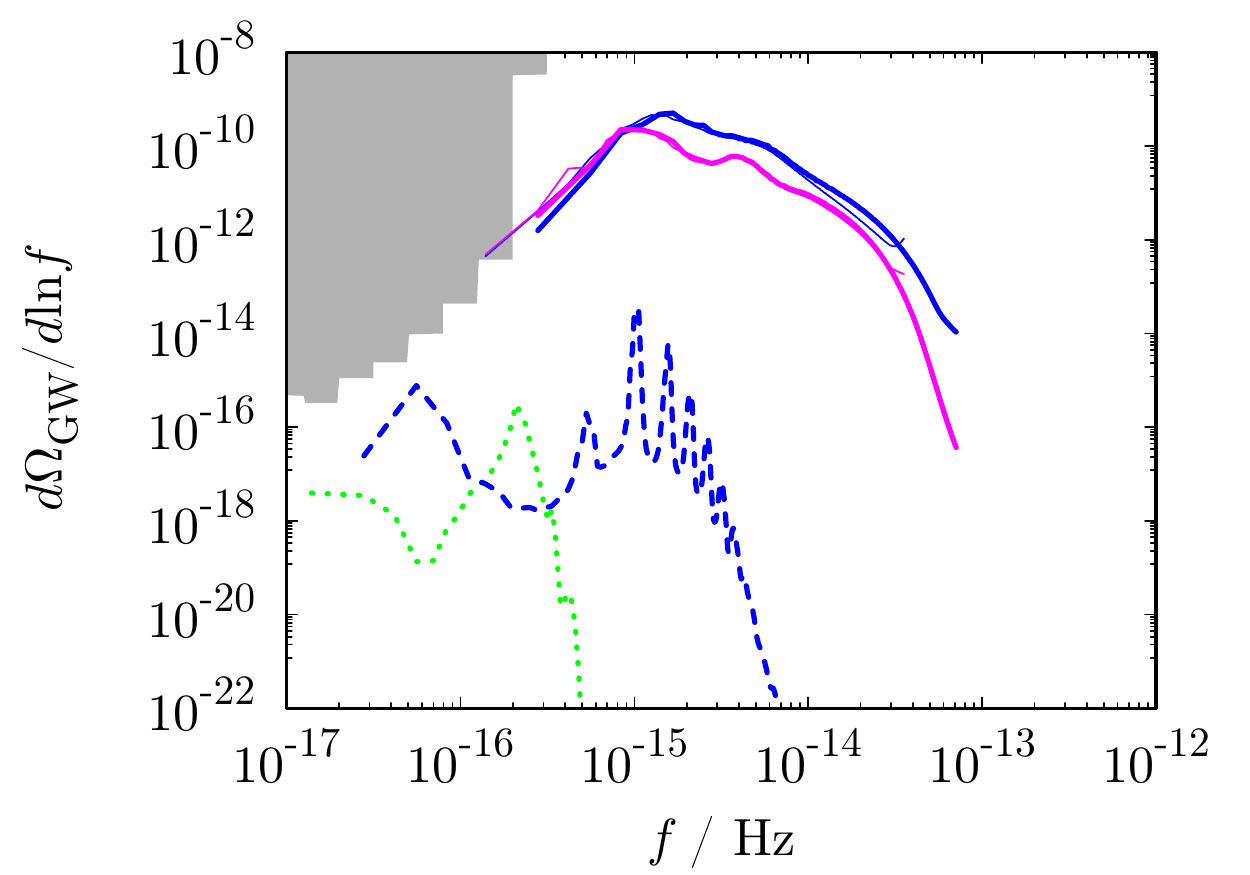}
\caption{
Spectrum of the GW density parameter at $z=1000$ for cases (c) and (d) (solid lines), $100$ for case (b) (dashed line),
and $10$ for case (a) (dotted line). We have taken 
the initial value $\phi_i/f =\phi_i /\sqrt{6\alpha}M_P = 3$. Thin lines show the results with larger box size ($0.08\pi H_{\rm eq}^{-1}$ with case (c) and (d)) and the gray shaded region represents the current constraint from the CMB observation \cite{Clarke:2020bil}.
}
\label{fig:OmegaGW_fin}
\end{figure}

As shown in the previous section, fluctuations of the scalar EDE field can be exponentially amplified as a consequence of the tachyonic  and/or the parametric resonance instabilities.
Such a violent process leads to the gravitational wave emission.
The gravitational wave is defined by the tensor metric perturbation $h_{ij}$ on the background FLRW metric which is given by
\begin{align}
ds^2 = dt^2-a^2(t)(\delta_{ij}+h_{ij})dx^i dx^j.
\end{align}
Here and in what follows, the sum is taken over repeated indices unless otherwise stated.
The evolution of gravitational waves is derived from the linearlized Einstein equation: 
\begin{align} \label{eq:GWeom}
\ddot{h}_{ij}+3H\dot{h}_{ij}-\frac{\nabla^2 h_{ij}}{a^2} = -16\pi G \Lambda^{kl}_{ij} \Pi_{kl}
\end{align}
where $\Lambda^{kl}_{ij}$ is the transverse-traceless projection tensor, $\Pi_{ij} = -\partial_i \phi \partial_j \phi /a^2$ is the anisotropic stress originated from 
the scalar field and $G$ is the Newton's constant.
The energy density of stochastic GW is given by
\begin{align}
\rho_{\rm GW}(t) =  \frac{1}{32\pi G} \langle \dot{h}_{ij}(\bm{x}) \dot{h}_{ij} (\bm{x}) \rangle,
\end{align}
where the angular bracket represents the ensemble average. Practically, this average can be replaced with the spatial average in the lattice simulation. The spectrum of GW density parameter is defined by
\begin{align}
\Omega_{\rm GW}(f) = \frac{1}{\rho_{\rm cr}} \frac{d\rho_{\rm GW}}{d \ln f},
\end{align}
where $\rho_{\rm cr} = 3H^2 M_P^2$ is the critical density and $f$ is the frequency of GW. 

We have solved the evolution equation for GWs (\ref{eq:GWeom}) together with the scalar field evolution (\ref{eq:scalarEoM}) with the numerical lattice simulation following \cite{Garcia-Bellido:2007fiu}. Here we are interested in the induced GWs, and thus we neglect the primordial GW background from the inflationary origin and set $h_{ij}=0$ as the initial condition for our simulation. 

The time evolution of the GW density parameter is shown in Fig.~\ref{fig:OmegaGW} for each case.
In the axion EDE case (figure \ref{subfig:OmegaGW_a}), the resonant amplification is not efficient and thus the resultant GW amplitude (at $z=1$ shown by thick black line) is smaller than that in other cases. 
The final GW amplitude is much larger in a simple $\alpha$-attractor model (figure \ref{subfig:OmegaGW_b}),  but still most GW emission occurs only after the recombination epoch. Note that, in these cases, the multiple-peak structure appears as an imprint of the narrow resonance band structure. 
However, the emitted GWs cannot be probed by CMB in those cases.
In the cases (c) and (d) (figure \ref{subfig:OmegaGW_c} and \ref{subfig:OmegaGW_d}), low-$k$ modes are efficiently excited due to the tachyonic instability,
which induces large scalar fluctuations.
The amplitude of the emitted GW is many orders of magnitude larger than the case with top panels\footnote{
The density parameter of GWs looks larger than 1 in this figure, but 
notice that the vertical axis is normalized by $(f/M_P)^4 \sim 10^{-4}$, and hence $\Omega_{\rm GW}$ appears to be big, however $\Omega_{\rm GW}$ does not exceed unity.
}. In this case, the efficient GW emission occurs before the last scattering surface and it can affect the CMB observation.

Figure~\ref{fig:OmegaGW_fin} shows the resultant GW spectrum at $z=1000$ (solid line) for the cases (c) and (d), 100 (dashed line) for the case (b), and 10 (dotted line) for the case (a)\footnote{At that time, the GW production is not yet completed and thus the multiple-peak structure does not yet appear in the GW spectrum.}.
Although, due to the limitation of the lattice simulation, we have not evaluated the GW spectrum at lower frequency, especially for the cases (c) and (d), however
if one simply extrapolates the solid lines to lower frequency, e.g. $10^{-16}$ Hz, the CMB constraints may easily rule out the model \cite{Clarke:2020bil}.

\section{Summary and discussion} \label{sec:disc}

In this paper, we have studied the GW emission from the EDE field fluctuations amplified by the tachyonic/parametric resonance instabilities due to the self-interactions. 
We have shown by performing the numerical lattice simulations that the tachyonic instability efficiently enhances the field fluctuations and the GWs 
can be produced 
before the recombination epoch
while the narrow resonance is less efficient and the GW emission is delayed. Note that the tachyonic growth occurs near the minimum of the potential regardless of 
the power index of the potential for the generalized $\alpha$-attractor type. 

As discussed in Section~\ref{sec:res}, 
the nonlinear evolution of EDE may have a great impact not only on the production of GWs, but also on 
the prospect of EDE as a possible solution of the Hubble tension.
In many works which study the fit of the EDE to CMB and some other data, the evolutions of the background and linear perturbation have been implemented in the analysis.
However, we have shown that the scalar field fluctuation can dominate over the coherent oscillation. In such a case, the equation of state of the EDE can be modified as in the preheating case \cite{Podolsky:2005bw}. Indeed, as shown in Fig.~\ref{fig:evolve}, the nonlinear effect significantly reduces the energy density of the EDE compared with the homogeneous case. As a consequence, it possibly changes the fit to CMB and other cosmological data. Thus, one should correctly take into account the temporal change of the equation of state of the EDE in the analysis to investigate its constraints in some cases\footnote{
The prediction for the cosmic birefringence can also be affected. In particular, since the axion EDE field becomes highly inhomogeneous, the anisotropic birefringence is expected rather than the isotropic one and it can also be constrained by the CMB data \cite{Gruppuso:2020kfy,Bortolami:2022whx}.
}.

In this paper, we simply assume that the EDE does not have interaction with other fields.
As an extension of this work, one can consider a multi-field model allowing the interaction between the EDE and other spectator fields (as in the new EDE model \cite{Niedermann:2019olb,Niedermann:2020dwg,Cruz:2023cxy}).  In such a case, the resonant production of spectator fields can be efficient and it can enhance the GW emission accordingly. 
This issue would also be worth investigating, which is left for future work.

\section*{Acknowledgment}
This work was supported by JSPS KAKENHI Grant Number 19K03874 (TT), 20H01894 (NK), 20H05851 (NK), 21H01078 (NK), 21KK0050 (NK) and MEXT KAKENHI 23H04515 (TT).

\bibliographystyle{utphys}
\bibliography{ref}

\end{document}